# Ring-shaped potential and a class of relevant integrals involved universal associated Legendre polynomials with complicated arguments

Wei Li[1,a)], Chang-Yuan Chen[2,b)], Shi-Hai Dong[3]

[1)]School of Science, Beijing University of Chemical Technology, Beijing, 100029, P. R. China

[2)]New energy and electronics, Yancheng Teachers University, Yancheng, 224002, P. R. China

[3)]CIDETEC, Instituto Politécnico Nacional, Unidad Professional ALM, Ciudad de México 07700,Mexico

## Abstract

We find that the solution of the polar angular differential equation can be written as the universal associated Legendre polynomials. Its generating function is applied to obtain an analytical result for a class of interesting integrals involving complicated argument i.e. $\int_{-1}^{1} P_{l'}^{m'}\left(\frac{xt-1}{\sqrt{1+t^2-2xt}}\right) \frac{P_{k'}^{m'}(x)}{(1+t^2-2tx)^{(l'+1)/2}} dx$ where $t \in (0,1)$. The present method can in principle be generalizable to the integrals involving other special functions. As an illustration we also study a typical Bessel integral with a complicated argument $\int_0^\infty \frac{J_n(\alpha\sqrt{x^2+z^2})}{\sqrt{(x^2+z^2)^n}} x^{2m+1} dx$.

It is well known that the associated Legendre polynomials play an important role in the central fields when one solves the physical problems in the spherical coordinates. However, in the case of the non-central fields we have to introduce the universal associated Legendre polynomials $P_{l'}^{m'}(x)$ when one studies the modified Pöschl-Teller [1], the single and double ring-shaped potentials and time-dependent potential [2-4]. Among them, the single ring-shaped potential can be written as the following form

$$V(r,\theta) = V(r) + \frac{\hbar^2}{2M\, r^2}\frac{b}{\sin^2\theta}, \qquad (1)$$

where the $V(r)$ might be taken as the Coulomb potential or the spherical harmonic oscillator. In spherical coordinates, by taking $\Psi(\vec{r}) = r^{-1}u(r)H(\theta)\mathrm{e}^{\pm im\varphi}$, $m = 0,1,2,\cdots$,

---

[a] E-mail address: liwei2@mail.buct.edu.cn
[b] Corresponding authors: C. Y. Chen (E-mail: yctcccy@163.net) and S.H. Dong (E-mail: dongsh2@yahoo.com)

and then substituting it into the Schrödinger equation

$$\left(-\frac{\hbar^2}{2M}\nabla^2 + V(r,\theta)\right)\Psi(\vec{r}) = E\Psi(\vec{r}), \qquad (2)$$

we obtain the polar angular differential equation

$$\frac{1}{\sin\theta}\frac{d}{d\theta}\left(\sin\theta\frac{dH(\theta)}{d\theta}\right) + \left(\lambda - \frac{m^2+b}{\sin^2\theta}\right)H(\theta) = 0. \qquad (3)$$

If taking $x = \cos\theta \in [-1,1]$, then equation (3) is modified as a universal associated-Legendre differential equation

$$(1-x^2)\frac{d^2H(x)}{dx^2} - 2x\frac{dH(x)}{dx} + \left[l'(l'+1) - \frac{(m')^2}{1-x^2}\right]H(x) = 0, \qquad (4)$$

where $\lambda = l'(l'+1)$, $m' = \sqrt{b+m^2}$. The choice of positive or negative $b$ depends on $b \geq -m^2$. Note that the quantum numbers $l'$ and $m'$ are not taken as integer. The series expression of the solution is given by

$$P_{l'}^{m'}(x) = (1-x^2)^{m'/2} \sum_{\nu=0}^{\left[\frac{l'-m'}{2}\right]} \frac{(-1)^\nu \Gamma(2l'-2\nu+1)}{2^{l'}\nu!(l'-m'-2\nu)!\Gamma(l'-\nu+1)} x^{l'-m'-2\nu}, \qquad (5)$$

with the following properties

$$\int_{-1}^{+1}[P_{l'}^{m'}(x)]^2 dx = \frac{2\Gamma(l'+m'+1)}{(2l'+1)n!}, \quad \int_{-1}^{+1}\frac{[P_{l'}^{m'}(x)]^2}{1-x^2}dx = \frac{\Gamma(l'+m'+1)}{m'n!}. \qquad (6)$$

Just recently, we have studied some useful integrals involving the product of the universal associated Legendre polynomials with some simple arguments, say $x$ [5]. These studies were stimulated by the fact that we have to calculate the following integrals $I_A^\pm(a,\tau) = \int_{-1}^{+1} x^a [P_{l'}^{m'}(x)]^2 / (1\pm x)^\tau dx$, $I_B(b,\sigma) = \int_{-1}^{+1} x^b [P_{l'}^{m'}(x)]^2 / (1-x^2)^\sigma dx$ and $I_C^\pm(c,\kappa) = \int_{-1}^{+1} x^c [P_{l'}^{m'}(x)]^2 / (1-x^2)^\kappa (1\pm x) dx$ when the $P_{l'}^{m'}(x)$ is applied to some physical problems, e.g., in studying the spin-orbit interaction for the ring-shaped potential [6] and matrix elements in relevant topics. These integrals cannot be found in classical integral handbooks [7, 8]. On the other hand, the parameters $a,b,c,\tau,\sigma,\kappa$ appearing in above type of integrals cannot be taken arbitrary values because of the complexity of integrals [5]. To attack this problem, we have used the properties of the generalized hypergeometric functions to calculate those integrals for arbitrary parameter values [9] since these definite integrals become significant both in mathematics and in theoretical physics. It should be pointed out that the argument appearing in the polynomial $P_{l'}^{m'}(x)$ involved in these integrals mentioned above [5, 6, 9] is simple, i.e., the variable $x$. However, sometimes the complicated argument will be occurred in the calculations as shown in classical handbooks. Since the universal associated Legendre polynomials are the generalized associated Legendre polynomials. The latter has wide applications in central fields, in particular in high energy physics and in nuclear physics, etc.

In this work, our main purpose is to attack a class of integrals of universal associated Legendre polynomials with complicated arguments through the generating functions since such an integral is helpful in high energy physics and in nuclear physics. The charm of the generating functions lies in making the complicated integral easy and soluble. The integral also includes another variable $t \in (0,1)$ except for the complicated argument and usual integral variable $x$. That is, the integrals that we are going to study are expressed as

$$I = \int_{-1}^{1} P_{l'}^{m'}\left(\frac{xt-1}{\sqrt{1+t^2-2xt}}\right) \frac{P_{k'}^{m'}(x)}{(1+t^2-2tx)^{(l'+1)/2}} dx. \qquad (7)$$

Moreover, we generalize the present approach to study the integral for the Bessel function with a complicated argument.

Let us study the first integral (7). As shown in our recent work [10], the generating function of the universal associated Legendre polynomials is given by

$$g(x,v) = \frac{\Gamma(2m'+1)}{2^{m'}\Gamma(m'+1)}(1-x^2)^{m'/2}\left(1-2xv+v^2\right)^{-m'-1/2} = \sum_{n=0}^{\infty} P_{l'}^{m'}(x)v^n, \quad |v|<1, \qquad (8)$$

where $n = l'-m'$. In particular, when $l' = l, m' = m$, the $g(x,v)$ reduces to the case of the well-known associated Legendre polynomials. For convenience, if changing the running number $n = l'-m'$, this generating function can also be modified as

$$g(x,v) = \frac{\Gamma(2m'+1)}{2^{m'}\Gamma(m'+1)}(1-x^2)^{m'/2}\left(1-2xv+v^2\right)^{-m'-1/2} v^{m'} = \sum_{l'=m'}^{\infty} P_{l'}^{m'}(x)v^{l'}. \qquad (9)$$

Let

$$v \to \frac{u}{\sqrt{1+t^2-2tx}}, \qquad x \to \frac{xt-1}{\sqrt{1+t^2-2tx}}. \qquad (10)$$

Considering the generating function (9), one may write out the following expression

$$\frac{1}{(1+t^2-2tx)^{1/2}} P_{l'}^{m'}\left(\frac{xt-1}{\sqrt{1+t^2-2tx}}\right) v^{l'}$$

as follows:

$$\frac{\left[1-\frac{(xt-1)^2}{(1+t^2-2tx)}\right]^{m'/2}}{\sqrt{1+t^2-2tx}} \frac{\Gamma(2m'+1)}{2^{m'}\Gamma(m'+1)} \frac{u^{m'}}{(1+t^2-2tx)^{m'/2}} \left[1+\left(\frac{u}{\sqrt{1+t^2-2tx}}\right)^2 - \frac{2u(xt-1)}{\left(\sqrt{1+t^2-2tx}\right)^2}\right]^{-(2m'+1)/2} \quad (11)$$

$$= \sum_{l'=m'}^{\infty} \left[ P_{l'}^{m'}\left(\frac{xt-1}{\sqrt{1+t^2-2tx}}\right)(1+t^2-2tx)^{-(l'+1)/2} u^{l'} \right].$$

The left hand side of Eq. (11) can be simplified as

$$\frac{t^{m'}(1-x^2)^{m'/2} u^{m'}}{\left[(1+u)^2+t^2-2(1+u)tx\right]^{m'+1/2}} \frac{\Gamma(2m'+1)}{2^{m'}\Gamma(m'+1)} = \frac{u^{m'}}{(1+u)^{m'+1}} \frac{\Gamma(2m'+1)}{2^{m'}\Gamma(m'+1)} \frac{(1-x^2)^{m'/2} s^{m'}}{(1+s^2-2sx)^{m'+1/2}}, \quad (12)$$

where we have used the notation $s \equiv t/(1+u)$ and the following identities

$$\frac{1}{\sqrt{(1+u)^2+t^2-2tx(u+1)}} = \frac{1}{(1+u)} \frac{1}{\sqrt{1+s^2-2sx}}. \quad (13)$$

In this case, the right hand side of Eq.(11) can be written as

$$\sum_{l'=m'}^{\infty} \left[ P_{l'}^{m'}\left(\frac{xt-1}{\sqrt{1+t^2-2tx}}\right)(1+t^2-2tx)^{-(l'+1)/2} u^{l'} \right] = \frac{u^{m'}}{(1+u)^{m'+1}} \frac{\Gamma(2m'+1)}{2^{m'}\Gamma(m'+1)} \frac{(1-x^2)^{m'/2} s^{m'}}{(1+s^2-2sx)^{m'+1/2}}. \quad (14)$$

Thus, based on above formula and Eq. (9) and by integrating it we have

$$\sum_{l'=m'}^{\infty} \sum_{k'=m'}^{\infty} u^{l'} v^{k'} \int_{-1}^{+1} P_{l'}^{m'}\left(\frac{xt-1}{\sqrt{1+t^2-2tx}}\right) \frac{P_{k'}^{m'}(x)}{(1+t^2-2tx)^{(l'+1)/2}} dx$$

$$= \frac{u^{m'}}{(1+u)^{m'+1}} \int_{-1}^{+1} \left( \frac{\Gamma(2m'+1)}{2^{m'}\Gamma(m'+1)} \frac{(1-x^2)^{m'/2} s^{m'}}{(1+s^2-2sx)^{m'+1/2}} \right) \left( \frac{\Gamma(2m'+1)}{2^{m'}\Gamma(m'+1)} \frac{(1-x^2)^{m'/2} v^{m'}}{(1+v^2-2vx)^{m'+1/2}} \right) dx. \quad (15)$$

It is found from the right side hand of identity (15) that each of the two factors of the integrand is the generating function of the *m*-th universal associated Legendre polynomials. By expanding each of them into the sum of the *m*-th universal associated Legendre polynomials and then using its orthogonal relation

$$\int_{-1}^{1} P_{l'}^{m'}(x) P_{k'}^{m'}(x) dx = \frac{2}{2l'+1} \frac{\Gamma(l'+m'+1)}{\Gamma(l'-m'+1)} \delta_{l'k'}, \quad (16)$$

one can evaluate the right hand side of Eq. (12) to yield

$$\frac{u^{m'}}{(1+u)^{m'+1}} \sum_{k'=m'}^{\infty} \frac{2}{2k'+1} \frac{\Gamma(k'+m'+1)}{\Gamma(k'-m'+1)} v^{k'} s^{k'}. \quad (17)$$

The coefficient of the factor $v^{k'} s^{k'}$ term must come from the coefficient of the $u^{l'}$ of the following expansion

$$\frac{2}{(2k'+1)} \frac{u^{m'} s^{k'}}{(1+u)^{m'+1}} = \frac{2}{(2k'+1)} \frac{t^{k'} u^{m'}}{(1+u)^{k'+m'+1}} = \frac{2t^{k'}}{(2k'+1)} \sum_{n=0}^{\infty} (-1)^n \frac{\Gamma(k'+m'+n+1)}{\Gamma(n+1)\Gamma(k'+m'+1)} u^{m'+n}$$

$$= \frac{2t^{k'}}{(2k'+1)} \sum_{l'=m'}^{\infty} (-1)^{l'-m'} \frac{\Gamma(k'+l'+1)}{\Gamma(l'-m'+1)\Gamma(k'+m'+1)} u^{l'}. \quad (18)$$

Therefore, the integral on the right hand side of Eq. (15) can be simplified as

$$\sum_{l'=m'}^{\infty}\sum_{k'=m'}^{\infty} u^{l'} v^{k'} \frac{2t^{k'}}{2k'+1} \frac{(-1)^{l'-m'}\Gamma(l'+k'+1)}{\Gamma(l'-m'+1)\Gamma(k'-m'+1)} \quad . \tag{19}$$

Finally, we obtain the following useful integral formula

$$I = \int_{-1}^{1} P_{l'}^{m'}\left(\frac{xt-1}{\sqrt{1+t^2-2xt}}\right) \frac{P_{k'}^{m'}(x)}{(1+t^2-2tx)^{(l'+1)/2}} dx$$
$$= \frac{2t^{k'}}{2k'+1} \frac{(-1)^{l'-m'}\Gamma(k'+l'+1)}{\Gamma(l'-m'+1)\Gamma(k'-m'+1)} \quad . \tag{20}$$

We are now in the position to generalize the present approach to other special functions. As an illustration, we show how this method can be applied to solve the following integral [7]

$$\int_0^{\infty} \frac{J_n(\alpha\sqrt{x^2+z^2})}{\sqrt{(x^2+z^2)^n}} x^{2m+1} dx = \frac{2^m \Gamma(m+1)}{\alpha^{m+1} z^{n-m-1}} J_{n-m-1}(\alpha z), \tag{21}$$

with $\alpha > 0$ and $\operatorname{Re} m > -1$ to make this integral convergent. We begin by the generating functions of the Bessel functions, which is given by [11,12]

$$e^{x(t-1/t)/2} = \sum_{n=-\infty}^{\infty} J_n(x) t^n \quad . \tag{22}$$

With the aid of this, it is not difficult to write out the following expressions

$$e^{\frac{\alpha\sqrt{x^2+z^2}}{2}\left(\frac{t}{\sqrt{x^2+z^2}} - \frac{\sqrt{x^2+z^2}}{t}\right)} = \sum_{n=-\infty}^{\infty} \frac{J_n(\alpha\sqrt{x^2+z^2})}{\sqrt{(x^2+z^2)^n}} t^n \quad . \tag{23}$$

Let us calculate the following integral

$$\int_0^{\infty} x^{2m+1} e^{\frac{\alpha\sqrt{x^2+z^2}}{2}\left(\frac{t}{\sqrt{x^2+z^2}} - \frac{\sqrt{x^2+z^2}}{t}\right)} dx = \exp\left[\frac{\alpha t}{2} - \frac{\alpha z^2}{2t}\right] \int_0^{\infty} x^{2m+1} e^{-\frac{\alpha}{2t}x^2} dx$$
$$= \exp\left[\frac{\alpha t}{2} - \frac{\alpha z^2}{2t}\right] \frac{2^m \Gamma(m+1)}{\alpha^{m+1}} t^{m+1}, \quad m > -1, \tag{24}$$

where we have used the formula [7]

$$\int_0^{\infty} x^m e^{-\beta x^n} dx = \frac{\Gamma(\gamma)}{n\beta^{\gamma}}, \quad \gamma = \frac{m+1}{n}, \quad \operatorname{Re}\beta > 0, \operatorname{Re} m > 0, \operatorname{Re} n > 0. \tag{25}$$

After some algebraic manipulations, it is not difficult to see that

$$\exp\left[\frac{\alpha t}{2} - \frac{\alpha z^2}{2t}\right] \frac{2^m \Gamma(m+1)}{\alpha^{m+1}} t^{m+1} = \exp\left[\frac{\alpha z}{2}\left(\frac{t}{z} - \frac{z}{t}\right)\right] \frac{2^m \Gamma(m+1)}{\alpha^{m+1}} t^{m+1}$$
$$= \frac{2^m \Gamma(m+1)}{\alpha^{m+1}} \sum_s J_s(\alpha z) \frac{t^s}{z^s} t^{m+1}. \tag{26}$$

The coefficient of the $t^n$ in the above expression is given by choosing $s = n-m-1$. As a result, we can get the final result (21).

In this paper, we have evaluated a class of integral for universal associated Legendre polynomials with complicated arguments through using the generating functions. Also, we generalized this method to derive the integral of the Bessel function with a complicated argument. Undoubtedly, this provides an effective approach for dealing with those integrals with complicated arguments.

**Acknowledgement**: We would like to thank the referees for invaluable and positive suggestions which have improved the manuscript greatly. This work is supported by the National Natural Science Foundation of China under Grant No. 11275165 and partially by 20170938-SIP-IPN, Mexico.

**Competing Interests**: The authors declare that there is no conflict of interest regarding the publication of this paper.

Press: Boston, New York, 2005.

[12] M. Abramowitz, I. A. Stegun, *Handbook of Mathematical Functions: with Formulas, Graphs, and Mathematical Tables,* Dover Publications, 1965.
7